\documentclass[prb,amsfonts,amssymb,floats,twocolumn,aps]{revtex4}

\usepackage[final,dvips]{epsfig}
\usepackage{bm}

\begin{document} 

\title[Two-site dynamical mean-field theory]
      {Two-site dynamical mean-field theory}

\author{Michael Potthoff}

\email[email: ]{michael.potthoff@physik.hu-berlin.de}

\affiliation{
Institut f\"ur Physik, 
Humboldt-Universit\"at zu Berlin, 
10115 Berlin, 
Germany
}
 
\begin{abstract}
It is shown that a minimum realization of the dynamical mean-field theory (DMFT) 
can be achieved by mapping a correlated lattice model onto an impurity model in 
which the impurity is coupled to an uncorrelated bath that consists of a single 
site only.
The two-site impurity model can be solved exactly. 
The mapping is approximate.
The self-consistency conditions are constructed in a way that the resulting 
``two-site DMFT'' reduces to the previously discussed linearized DMFT for 
the Mott transition.
It is demonstrated that a reasonable description of the mean-field physics 
is possible with a minimum computational effort. 
This qualifies the simple two-site DMFT for a systematic study of more complex 
lattice models which cannot be treated by the full DMFT in a feasible way.
To show the strengths and limitations of the new approach, the single-band
Hubbard model is investigated in detail.
The predictions of the two-site DMFT are compared with results of the full DMFT.
Internal consistency checks are performed which concern the Luttinger sum rule, 
other Fermi-liquid relations and thermodynamic consistency.
\end{abstract} 
 

 
\maketitle 

\section{Introduction}

The dynamical mean-field theory (DMFT) \cite{GK92a,Jar92,PJF95,GKKR96} has become 
a well-established and valuable method to investigate the physics of strongly 
correlated electrons on a lattice.
Similar as the Weiss mean-field theory for classical models of localized
spins, the DMFT is exact in the non-trivial limit of infinite spatial 
dimensions. \cite{MV89,Vol89}
For finite dimensions it provides a thermodynamically consistent and 
non-perturbative mean-field approach in which the spatial correlations are
neglected but the temporal degrees of freedom are treated correctly. 
The application of the DMFT to the single-band Hubbard model 
\cite{Hub63,Gut63,Kan63} has uncovered a complex phenomenology 
which may be characterized by strongly renormalized Fermi-liquid behavior competing
with the Mott insulating state and with different kinds of spontaneous order 
such as collective magnetism. \cite{PJF95,GKKR96,Ulm98,VBH+97,OPK97}

The DMFT actually consists in a prescription that maps the original
lattice model onto an effective impurity model which describes a single 
correlated impurity orbital 
embedded in an uncorrelated bath of conduction-band states.
This mapping is a self-consistent one, namely the bath parameters depend on
the on-site lattice Green function.
The impurity model is the crucial point in the DMFT since it poses a highly
non-trivial many-body problem that must be solved repeatedly.
The different methods employed for an essentially exact solution of the 
impurity model, the quantum Monte-Carlo, \cite{Jar92,RZK92}
the exact diagonalization \cite{CK94,SRKR94}
and the numerical renormalization-group method, \cite{BHP98,Bul99}
work well for the single-band Hubbard
model but are computationally expensive.
In practice this severely limits the applicability of the DMFT, particularly
when one is concerned with multi-band models.
More complex lattice models including two, three or more degenerate or 
non-degenerate $d$-like bands possibly hybridized with uncorrelated $s$-$p$-like 
bands are interesting for obvious reasons. 
For example, multi-band models are required for a minimum theory of the physics 
of strongly correlated electrons in the transition-metal oxides.
Because of the large parameter space and the complexity of the effective impurity
problem, a detailed and systematic calculation of the phase diagrams of 
multi-band models covering the entire parameter space is far beyond the 
ability of present implementations of the DMFT.

If one is interested in a comprehensive mean-field description of complex lattice 
models but wishes to keep the essence of the DMFT, some compromise is inevitable.
In principle, there are only two possibilities conceivable: 
First, one may refrain from a numerically exact solution of the impurity
problem and employ approximate treatments instead. 
These may be based on different limits where a small parameter is available. 
This idea has been pursued with weak- and strong-coupling perturbational approaches 
such as the iterative perturbation theory \cite{GK92a,KK96,PWN97} or the 
non-crossing approximation. \cite{PCJ93,HS94,OPK97}
In fact, multi-band Hubbard-type models can be treated in this way with an
acceptable computing time (see Refs.\ \onlinecite{LK98,MZPK99}, for examples).
This route, however, shall not be followed up here.

The present paper takes into consideration the only alternative left,
namely to solve the impurity model exactly but to reduce the number of bath 
degrees of freedom to keep the calculations manageable.
There are no problems to treat e.\ g.\ the single-impurity Anderson 
model (SIAM) with a small number of sites $n_{\rm s}$ numerically exact.
On the other hand, for any finite $n_{\rm s} < \infty$ the self-consistent 
mapping of the Hubbard model onto the SIAM is approximate. 
The exact solution of the effective impurity model is thus achieved at the 
expense of an approximate self-consistency.

As a function of $n_{\rm s}$ the Hilbert-space dimension $D$ of the impurity
model increases exponentially.
It is given by $D = 2^{2M}$ where $M$ is the number of (two-fold spin-degenerate)
one-particle orbitals.
The self-consistent mapping within the DMFT at least requires 
$M = r + r (n_{\rm s}-1) = r n_{\rm s}$ orbitals for the case of one impurity 
site, $n_{\rm s}-1$ bath sites, and $r$ correlated bands (see appendix).
Consider, for example, the $d$-band of a $3d$ transition metal with a 
two-fold degenerate $e_g$-derived band and a three-fold degenerate 
$t_{2g}$-derived band. 
In this case $M = 5 n_{\rm s}$ orbitals have to be considered.
Accepting $M=10$ as a typical value for the 
maximum number of orbitals that can be taken into account for a repeated 
solution of the impurity model within a reasonable computing time, limits 
$n_{\rm s}$ to the smallest number that is reasonable: $n_{\rm s}=2$, 
i.\ e.\ an impurity model with one impurity site and one bath site.

The purpose of the present paper is to test whether or not a two-site DMFT 
could be a meaningful approach and to show up its strengths and limitations.
There are mainly two tasks to be performed: 
First, it is necessary to specify the approximate self-consistent mapping.
This means to find a sensible prescription how to fix the bath parameters
of the impurity model -- guided by the original self-consistency condition 
of the DMFT.
Second, the resulting two-site DMFT has to be tested against the full DMFT.
This can be done best for the single-band Hubbard model (in infinite dimensions)
where numerous essentially exact results are available.
Thus, except for a short discussion of the extension of the theory to multi-band 
systems, the present paper is exclusively concerned with the single-band model.
Since one cannot expect a strongly simplified two-site approach to reproduce 
the known results quantitatively exact, the present study attaches importance 
to main trends, qualitative correctness and internal consistency.
Recall that even the full DMFT cannot be expected to give more than a qualitative 
(mean-field) description of the physics of transition-metal oxides.

The two-site SIAM has been considered within the context of the DMFT beforehand.
Lange \cite{Lan98} discussed the two-site SIAM at half-filling to investigate
renormalized versus unrenormalized perturbational approaches to the DMFT of
the Mott transition. 
However, a self-consistent mapping of the Hubbard model onto the two-site SIAM
was not considered. 
Bulla and Potthoff \cite{BP00} developed a linearized DMFT of the Mott transition.
The Hubbard model at half-filling and at the critical interaction strength 
$U=U_{\rm c}$ was self-consistently mapped onto the two-site SIAM resulting 
in a linear algebraic mean-field equation. 
The solution of the two-site SIAM is exact, the mapping is approximate.
It was found that, compared with the full DMFT, the linearized DMFT gives fairly 
good analytical estimates for $U_{\rm c}$ on different lattices.
The theory, however, is restricted to the critical point.
Further extensions of the linearized DMFT for the Mott transition have been 
developed for a periodic Anderson model by Held and Bulla \cite{HB00} and 
for a two-band Hubbard model by Ono et al. \cite{OBH00}
Extensions of the linearized DMFT to the Hubbard model for thin films \cite{PN99a} 
and semi-infinite lattices \cite{PN99d} have convincingly shown that 
the main trends in the geometry dependence of $U_{\rm c}$ can be predicted
safely. 
Here, it will be shown  that a two-site DMFT can be constructed which is not
bound to the (Mott) critical point but is able to access the entire parameter 
space and that reduces to the linearized DMFT at half-filling and $U=U_{\rm c}$. 

The paper is organized as follows:
The next section \ref{sec:2sdmft} gives a brief review of the DMFT and introduces
the self-consistent two-site approach. 
Section \ref{sec:results} presents a variety of results for the single-band Hubbard 
model starting with the Mott transition. 
Hereafter, the Fermi-liquid phase off half-filling is addressed, and the Luttinger 
sum rule, other Fermi-liquid relations and thermodynamical consistency are 
discussed. 
A discussion of the two-site DMFT in relation to other methods and the 
conclusions are given in section \ref{sec:conclusion}.
The generalization of the theory to multi-band models is presented in an 
appendix.

\section{Two-site dynamical mean-field theory}
\label{sec:2sdmft}

To be definite the single-band Hubbard model on the Bethe lattice with infinite 
connectivity $q \mapsto \infty$ is considered:
\begin{equation}
  H = \sum_{\langle ij \rangle \sigma} t_{ij} \: c_{i\sigma}^\dagger c_{j\sigma} 
  + \frac{U}{2} \sum_{i\sigma} n_{i\sigma} n_{i-\sigma} \: .
\label{eq:hubbard}
\end{equation}
The hopping is assumed to be non-zero between nearest neighbors $i$ and $j$ only.
$t \equiv -t_{ij}>0$ is the nearest-neighbor hopping integral.
The on-site hopping $t_0 \equiv t_{ii}$ is set to $t_0=0$ to fix the energy zero.
Furthermore, $U$ is the on-site Coulomb interaction,
$c^\dagger_{i\sigma}$ creates an electron at the site $i$ 
with spin $\sigma=\uparrow, \downarrow$, 
and $n_{i\sigma}=c^\dagger_{i\sigma}c_{i\sigma}$.
With the usual scaling of the hopping, $t=t^\ast/\sqrt{q}$ and
$t^\ast=\mbox{const}$, the model is non-trivial in the limit $q\mapsto \infty$.
\cite{MV89}
Setting $t^\ast=1$ fixes the energy scale for the present paper.
For a paramagnetic, spatially homogeneous phase the on-site Green function 
$G(\omega) = \langle\langle c_{i\sigma} ; c^\dagger_{i\sigma} 
\rangle\rangle_\omega$ is given by:
\begin{equation}
  G(\omega) = \int_{-\infty}^\infty dx
  \frac{\rho_0(x)}{\omega + \mu - x - \Sigma(\omega)} \: ,
\label{eq:green}
\end{equation}
where $\mu$ is the chemical potential, and $\Sigma(\omega)$ is the self-energy 
which is local (${\bf k}$ independent) in the limit $q \mapsto \infty$. 
\cite{MV89,MH89c}
$\rho_0(x)$ denotes the free density of states:
\begin{equation}
  \rho_{0}(x) = \frac{1}{2\pi {t^\ast}^2} \sqrt{4 {t^\ast}^2 - x^2} 
\label{eq:dos}
\end{equation}
for $|x|<2t^\ast$. The bandwidth is $W=4t^\ast=4$.

\subsection{Dynamical mean-field theory}

The DMFT essentially rests on the observation that the local self-energy
is given by a (skeleton-diagram) functional $\Sigma={\cal S}[G]$ of the on-site 
Green function that is universal for a large class of models.
Consider, in particular, the single-impurity Anderson model (SIAM):
\begin{eqnarray} 
  H_{\rm imp} &=& 
  \sum_{\sigma} \epsilon_{\rm d} d^\dagger_{\sigma} d_{\sigma} 
  + U d^\dagger_{\uparrow} d_{\uparrow} d^\dagger_{\downarrow} d_{\downarrow}
\nonumber \\
  &+& \sum_{\sigma, k=2}^{n_{\rm s}} 
  \epsilon_{k} a^\dagger_{k\sigma} a_{k\sigma}
  + \sum_{\sigma, k=2}^{n_{\rm s}}
  V_{k} (d^\dagger_\sigma a_{k\sigma} + \mbox{h.c.}) \: ,
\label{eq:imp} 
\end{eqnarray}
which describes an impurity orbital $d^\dagger_{\sigma}|0\rangle$ with one-particle 
energy $\epsilon_{\rm d}$ and on-site interaction $U$ that is coupled via the 
hybridization $V_k$ to a bath of $n_{\rm s}-1$ non-interacting orbitals 
$a^\dagger_{k\sigma} | 0 \rangle$ with energies $\epsilon_k$.
The impurity Green function 
$G_{\rm imp}(\omega)=\langle\langle d_{\sigma};
d^\dagger_{\sigma}\rangle\rangle_\omega$
is given by 
\begin{equation}
  G_{\rm imp}(\omega) =
  \frac{1}{\omega + \mu - \epsilon_{\rm d} 
  - \Delta(\omega) - \Sigma_{\rm imp}(\omega)} \: ,
\label{eq:impgreen}
\end{equation}
where $\Delta(\omega) = \sum_{k=2}^{\rm n_s} V_k^2 / (\omega + \mu - \epsilon_k)$
is the hybridization function, and $\Sigma_{\rm imp}(\omega)$ the impurity 
self-energy.
As usual, $\epsilon_{\rm d} = t_0 = 0$.
The important point is that the functional ${\cal S}$ is the same as for the 
Hubbard model, $\Sigma_{\rm imp}= {\cal S}[G_{\rm imp}]$, because the same
type of skeleton diagrams occur in the expansion of $\Sigma_{\rm imp}$.
Choosing the bath parameters $\epsilon_k$ and $V_k$ such that 
\begin{equation}
  \Delta(\omega) = \omega + \mu - \epsilon_{\rm d} - \Sigma_{\rm imp}(\omega)
  - \frac{1}{G(\omega)} \: ,
\label{eq:hybsc}  
\end{equation}
i.\ e.\ such that the DMFT self-consistency condition,
\begin{equation}
   G_{\rm imp}(\omega) \stackrel{!}{=} G(\omega) \: ,
\label{eq:sc}
\end{equation}
is fulfilled, then at once 
\begin{equation}
  \Sigma_{\rm imp}(\omega) = \Sigma(\omega) \: .
\label{eq:sigma}
\end{equation}

Therewith, the original lattice problem is mapped onto the SIAM and can be 
solved by the following iterative procedure: 
Starting with a guess for the local self-energy, the on-site lattice Green 
function is calculated from Eq.\ (\ref{eq:green}).
Via Eq.\ (\ref{eq:hybsc}) or Eq.\ (\ref{eq:sc}) the Green function and the 
self-energy define the hybridization function $\Delta(\omega)$ and thus 
the parameters of the effective SIAM. 
Finally, the impurity problem is solved to get a new estimate for the 
self-energy. The cycles have to be repeated until self-consistency is 
achieved.

\subsection{Two-site SIAM}

The self-consistency condition (\ref{eq:sc}) can be fulfilled rigorously only 
for $n_{\rm s} \mapsto \infty$, i.~e.\ for a bath with an infinite number of 
degrees of freedom. 
This leads to the usual SIAM which represents an involved many-body problem.
To simplify the problem and to construct a two-site DMFT, the case $n_{\rm s}=2$
is considered here, i.~e.\ an effective SIAM that consists of one impurity site
and one bath site only. 
This represents the most simple bath conceivable.

For $n_{\rm s}=2$ the site index is fixed to the value $k=2$ in Eq.\ (\ref{eq:imp}). 
There are only two independent bath parameters left, the one-particle energy of the 
bath site $\epsilon_{\rm c} \equiv \epsilon_{k=2}$ and the hybridization strength 
$V \equiv V_{k=2}$. 
The hybridization function is a one-pole function 
\begin{equation}
  \Delta(\omega) = V^2/(\omega + \mu - \epsilon_{\rm c}) \: ,
\end{equation}
and the free ($U=0$) impurity Green function is a two-pole function:
\begin{equation}
   G^{(0)}_{\rm imp}(\omega) = 
   \frac{1}{2r} \left(
   \frac{r + \overline{\epsilon}}{\omega+\mu-\overline{\epsilon} - r}
   +
   \frac{r - \overline{\epsilon}}{\omega+\mu-\overline{\epsilon} + r}
   \right)
   \: .
\label{eq:g0imp}
\end{equation}
with $\overline{\epsilon} \equiv (\epsilon_{\rm d} - \epsilon_{\rm c})/2$
and $r=\sqrt{\overline{\epsilon}^2+V^2}$.
The interacting impurity Green function $G_{\rm imp}(\omega)$ has four poles
and the self-energy $\Sigma_{\rm imp}(\omega)$ two poles in general.
Closed analytical expressions can be derived for the symmetric model at 
half-filling (see Ref.\ \onlinecite{Lan98}, for example).
For the non-symmetric case the model can be solved straightforwardly by numerical
means without any problems.

For clarity, the theory will be developed for the paramagnetic phase of single-band 
Hubbard model. 
However, it is rather straightforward to consider in essentially the same way 
also different magnetic phases and/or more complicated models such as multi-band 
Hubbard-type models, for example. 

\subsection{Self-consistency}

For the two-site DMFT the Eqs.\ (\ref{eq:green}) -- (\ref{eq:impgreen}) and
Eq.\ (\ref{eq:sigma}) are retained.
The original self-consistency condition (\ref{eq:sc}), however, must be 
reformulated.
This means to find two physically motivated (self-consistency) conditions to 
fix the bath parameters $\epsilon_{\rm c}$ and $V$.

Consider first the limit of high frequencies $\omega \mapsto\infty$. 
The exact self-energy of the impurity problem (\ref{eq:imp}) can be expanded 
in powers of $1/\omega$:
\begin{equation}
   \Sigma(\omega) = U n_{\rm d} + \frac{U^2 n_{\rm d} (1-n_{\rm d})}
   {\omega} + {\cal O}(1/\omega^{2}) \: ,
\label{eq:sighigh}
\end{equation}
where $n_{\rm d}=n_{\rm d\sigma}$ is the average occupancy of the impurity 
orbital:
\begin{equation}
   n_{\rm d} 
   = \langle d_\sigma^\dagger d_\sigma \rangle 
   = - \frac{1}{\pi} \int_{-\infty}^0 \mbox{Im} \: 
   G_{\rm imp}(\omega+i0^+) \: d\omega \: .
\label{eq:nimp}
\end{equation}
Inserting the expansion (\ref{eq:sighigh}) into (\ref{eq:green}), one
finds the following high-frequency expansion of the on-site lattice 
Green function:
\begin{eqnarray}
   G(\omega) &=& \frac{1}{\omega} 
             + \frac{t_0 - \mu + U n_{\rm d}}{\omega^2}
\nonumber \\
	     &+& \frac{M_2^{(0)} + (t_0-\mu)^2 + 2 (t_0 - \mu) U n_{\rm d} 
	     + U^2 n_{\rm d}} {\omega^3} 	     
\nonumber \\
	     &+& {\cal O}(1/\omega^{4}) \: .
\label{eq:greenhigh}
\end{eqnarray}
Here $M_2^{(0)} = \sum_{j \ne i}  t_{ij}^2 = \int dx \: x^2 \rho_0(x)$ is 
the variance of the non-interacting density of states (\ref{eq:dos}).
Eq.\ (\ref{eq:greenhigh}) can be compared with the exact high-frequency
expansion which is available by calculating the first non-trivial moments
of the interacting density of states. \cite{PHWN98}
One finds that Eq.\ (\ref{eq:greenhigh}) in fact represents the exact
expansion provided that $n_{\rm imp} \equiv 2n_{\rm d}$ can be identified 
with the filling
$n = \langle n_{i\uparrow} \rangle + \langle n_{i\downarrow} \rangle$ 
of the lattice model.
It is therefore required that
\begin{equation}
   n_{\rm imp} \stackrel{!}{=} n \: ,
\label{eq:cond1}
\end{equation}
where the band filling is calculated via
\begin{equation}
   n = - \frac{2}{\pi} \int_{-\infty}^0 \mbox{Im} \: 
   G(\omega+i0^+) \: d\omega \: .
\label{eq:n}
\end{equation}
Eq.\ (\ref{eq:cond1}) is the first self-consistency condition.
The high-frequency behavior of $G(\omega)$ is important for the occurrence 
and for the correct weights and centers of gravity of the two high-frequency 
Hubbard excitations in the spectrum.
With Eqs.\ (\ref{eq:nimp}), (\ref{eq:cond1}) and (\ref{eq:n}), an integral
form of the original self-consistency equation (\ref{eq:sc}) is fulfilled:
$\int^0_{-\infty} d\omega \, \mbox{Im} \, G 
=\int^0_{-\infty} d\omega \, \mbox{Im} \, G_{\rm imp}$.

Consider now the low-frequency limit $\omega \mapsto 0$.
The exact self-energy of the impurity problem (\ref{eq:imp}) can be 
expanded in powers of $\omega$, 
\begin{equation}
   \Sigma(\omega) = a + b \, \omega + {\cal O}(\omega^{2}) \: ,
\label{eq:siglow}
\end{equation}
with constants $a$ and $b$. 
The definition $z=1/(1-b)$ will be convenient, i.\ e.:
\begin{equation}
   z = \frac{1}{1-d\Sigma(0)/d\omega} \: .
\label{eq:zquasi}
\end{equation}
For a metal $z$ has the meaning of the quasi-particle weight.
Neglecting terms of the order $\omega^2$, and inserting into Eq.\ (\ref{eq:green}) 
yields $G(\omega) = G^{\rm (coh)}(\omega)$ for small $\omega$ where 
$G^{\rm (coh)}(\omega)$ is the coherent part of the on-site Green function 
defined as:
\begin{eqnarray}
  G^{\rm (coh)}(\omega) 
  &=& \int_{-\infty}^\infty dx \:
    \frac{\rho_0(x)}{\omega + \mu - x - a - b \omega} 
\nonumber \\
  &=& z \int_{-\infty}^\infty dx \:
    \frac{\rho_0(x)}{\omega - z(x - \mu + a)} \: .
\label{eq:greencoh}
\end{eqnarray}
On the other hand, the coherent part of the impurity Green function is:
\begin{eqnarray}
  G_{\rm imp}^{\rm (coh)}(\omega) 
  &=& \frac{1}{\omega + \mu - \epsilon_{\rm d} - \Delta(\omega) - a - b \omega} 
\nonumber \\
  &=& z \: \frac{1}{\omega - z (\epsilon_{\rm d} - \mu + a + \Delta(\omega))} \: .
\label{eq:greenimpcoh}
\end{eqnarray}
Comparing the {\em high}-frequency expansions of the respective coherent Green 
functions,
\begin{eqnarray}
  G^{\rm (coh)}(\omega) 
  &=& \frac{z}{\omega} + \frac{z^2 (t_0 - \mu + a)}{\omega^2} 
\nonumber \\
  &+& \frac{z^3 (M_2^{(0)} + (t_0 - \mu + a)^2)}{\omega^3} 
\nonumber \\
  &+& {\cal O}(1/\omega^4)
\label{eq:greencohhigh}
\end{eqnarray}
and
\begin{eqnarray}
  G_{\rm imp}^{\rm (coh)}(\omega) 
  &=& \frac{z}{\omega} + \frac{z^2 (\epsilon_{\rm d} - \mu + a)}{\omega^2} 
\nonumber \\
  &+& \frac{z^2 V^2 + z^3 (\epsilon_{\rm d} - \mu + a)^2}{\omega^3} 
\nonumber \\  
  &+& {\cal O}(1/\omega^4) \: ,
\label{eq:greenimpcohhigh}
\end{eqnarray}
leads to the second self-consistency condition:
\begin{equation}
  V^2 \stackrel{!}{=} z \: M_2^{(0)} \: .
\label{eq:cond2}
\end{equation}
Thereby, the original self-consistency equation (\ref{eq:sc}) is also fulfilled 
at low frequencies up to ${\cal O}(\omega)$ in an {\em integral} way, namely by 
referring to the weight, the center of gravity and the variance of the coherent 
quasi-particle peak. 
The equations (\ref{eq:cond1}) and (\ref{eq:cond2}) reformulate the high- and 
the low-frequency range of the original self-consistency condition (\ref{eq:sc})
in an integral, qualitative form and are thus well motivated.

\subsection{Calculations}

With the two conditions (\ref{eq:cond1}) and (\ref{eq:cond2}) the two bath 
parameters are fixed and can be calculated self-consistently:
Consider the model parameters $t_0, t, U, \mu$ and $\rho_{0}(x)$ to be given. 
Starting with a guess for $\epsilon_{\rm c}$ and $V$, the two-site impurity 
model (\ref{eq:imp}) is well defined and can be solved to find the average 
occupancy of the impurity level $n_{\rm imp}$ and the self-energy $\Sigma(\omega)$. 
The latter directly yields the quasi-particle weight $z=(1-d\Sigma(0)/d\omega)^{-1}$ 
and from the condition (\ref{eq:cond2}) a new value for the hybridization strength $V$.
Inserting $\Sigma(\omega)$ into Eq.\ (\ref{eq:green}), gives the on-site lattice
Green function $G(\omega)$ and, via Eq.\ (\ref{eq:n}), the filling $n$ which
has to be compared with $n_{\rm imp}$. 
Finally, a new value for $\epsilon_{\rm c}$ is chosen such that the difference
$n-n_{\rm imp}$ is reduced in the next cycle -- according to condition 
(\ref{eq:cond1}).
The cycles have to be iterated until $\epsilon_{\rm c}$ and $V$ are self-consistent
(and $n_{\rm imp}=n$).

In most cases, one finds a self-consistent set of bath parameters
$\epsilon_{\rm c}$ and $V$ such that the 
ground state of the effective SIAM lies in the 6 dimensional subspace characterized
by the total spin-dependent particle numbers $N_\uparrow = N_\downarrow = 1$.
To get the self-energy, it is convenient to calculate $G_{\rm imp}(\omega)$ 
from its Lehmann representation and to use the Dyson equation in reverse:
\begin{equation}
  \Sigma(\omega) = G^{(0)}_{\rm imp}(\omega)^{-1} - G_{\rm imp}(\omega)^{-1} \: ,
\end{equation}
where the free impurity Green function is taken from Eq.\ (\ref{eq:g0imp}).
The calculation of the eigenenergies and eigenstates is straightforward, but
even for $n_{\rm s}=2$ this can only be done numerically in general.
From the computational point of view, the most time-consuming step, however, 
is the calculation of the filling from Eqs.\ (\ref{eq:green}) and (\ref{eq:n}).
Since the self-energy of the two-site impurity problem is a real two-pole 
function of the form
\begin{equation}
   \Sigma(\omega) = \alpha_0 + \frac{\alpha_1}{\omega - \omega_1}
                             + \frac{\alpha_2}{\omega - \omega_2} \: ,
\label{eq:sigtwop}
\end{equation}
the filling can be calculated more directly by a single one-dimensional integration:
\begin{equation}
   n = 2 \int_{-\infty}^0 d\omega \: \rho_{0}(\omega + \mu - \Sigma(\omega)) \: .
\label{eq:fill}
\end{equation}
$\rho(\omega) \equiv \rho_{0}(\omega + \mu - \Sigma(\omega))$ is the
interacting density of states.
Since even a repeated evaluation of this equation during the search for
self-consistency is not very crucial, extremely fast numerical calculations
can be performed.

The two-site DMFT is obviously exact in the limits $n=0$ and $n=2$. 
In the empty-band limit, for example, the impurity self-energy vanishes since 
$n_{\rm imp} = n = 0$. 
Furthermore, it is exact in the band limit $U=0$.
In the atomic limit $t=0$ one finds $V=0$ (and arbitrary $\epsilon_{\rm c}$) 
to be a self-consistent solution. Since $n=n_{\rm imp}$, the self-energy is
given by 
\begin{equation}
  \Sigma_{\rm H-I}(\omega) = 
  U\frac{n}{2} + \frac{U^2(n/2)(1-n/2)}{\omega + \mu - t_0 - U(1-n/2)} \: ,
\label{eq:sighub}
\end{equation}
which is the correct atomic-limit self-energy. Hence, the two-site DMFT is 
also exact for $t=0$ and any filling. 
Since Eq.\ (\ref{eq:sighub}) is the Hubbard-I self-energy, \cite{Hub63} the 
theory reduces to the Hubbard-I approach whenever there is a self-consistent 
solution with $V=0$.
Actually, this is realized only at half-filling $n=1$ and for sufficiently 
strong $U$ (see below).

Although the original lattice problem is mapped onto an effective impurity 
problem with a finite number of degrees of freedom, the dependence of physical 
quantities, such as the filling $n$, on the model parameters will generally 
be smooth.
Consider, for example, the function $n(\mu)$ for a given $U$. The interacting
impurity spectral function generally consists of 4 isolated $\delta$ peaks of
different weight.
An infinitesimal change of $\mu$ is unlikely to cause a finite jump of 
$n_{\rm imp}$ since a change of $n_{\rm imp}$ is in first place caused by 
a redistribution of spectral weight among the $\delta$ peaks rather than by 
$\mu$ crossing a pole. 
Bearing in mind that the bath parameters itself depend on $\mu$, the function
$n(\mu)$ can be continuous even in a large $\mu$ interval.
In fact, it is found that the chemical potential never crosses a pole except
for some extreme cases ($n=0$, $n=2$, $V=0$).

\section{Results}
\label{sec:results}

The two-site DMFT provides a very simple, computationally fast and non-perturbative 
mean-field approach to correlated lattice models. 
To show its strengths and limitations, numerical results will be presented for 
the single-band Hubbard model which has been studied extensively by the full
DMFT in the past.
\cite{GK92a,Jar92,PJF95,GKKR96,MV89,Vol89,Ulm98,VBH+97,OPK97,RZK92,CK94,SRKR94,
BHP98,Bul99,KK96,PWN97,PCJ93,HS94,LK98,MZPK99}
This will set the basis for a discussion of the approach and a comparison with
other methods in section \ref{sec:conclusion}.
Multi-bands models are considered in the appendix.
The calculations have been performed for the Bethe lattice with infinite
connectivity $q$. The free density of states is given by Eq.\ (\ref{eq:dos}).
Its width is $W=4$. 

\subsection{Mott transition}

For the symmetric case of half-filling $n=1$ and $\mu=U/2$, the approach can 
be evaluated analytically. 
Particle-hole symmetry requires $\epsilon_{\rm c} = U/2 = \mu$ to ensure the
first self-consistency equation $n=n_{\rm imp}$.
Thus, only the hybridization strength $V$ has to be calculated self-consistently.

Let us first consider the critical point for the Mott transition, i.~e.\ 
$U \mapsto U_{\rm c}$ where the linearized DMFT is recovered 
(see Ref.\ \onlinecite{BP00}):
Coming from the weak-coupling, metallic side, the critical interaction 
$U_{\rm c}$ is characterized by a vanishing quasi-particle weight $z \mapsto 0$
which, according to Eq.\ (\ref{eq:cond2}) implies $V \mapsto 0$.
In this limit two of the four poles of the impurity Green function are located
near $\omega=\pm U/2$ while two poles lie close to $\omega = 0$. 
Referring to the latter,
the quasi-particle weight $z$ can be calculated as the sum of their weights. 
A straightforward calculation to leading order in $V$ yields:
\begin{equation}
   z = 2 \: \frac{18V^2}{U^2} \: .
\label{eq:smallv}
\end{equation}
On the other hand, another relation between $z$ and $V$ is given by the second
self-consistency condition (\ref{eq:cond2}).
This implies
\begin{equation}
   V^2 = M_2^{(0)} \: \frac{36}{U^2} \: V^2 \: ,
\end{equation}
namely a simple {\em linear} homogeneous mean-field equation as is characteristic 
for the linearized DMFT of Ref.\ \onlinecite{BP00}.
A non-trivial solution of this equation is only possible for $U=U_{\rm c}$ where
\begin{equation}
   U_{\rm c} = 6 \sqrt{M_2^{(0)}} \: .
\label{eq:uc}
\end{equation}
The second moment of $\rho_{0}(x)$ is easily calculated: 
$M_2^{(0)} = \sum_{j \ne i}  t_{ij}^2 = q t^2 = {t^\ast}^2 = 1$, and therewith
$U_{\rm c} = 6 t^\ast = 1.5 W$.
This result of the linearized DMFT is very close to the result of the 
projective self-consistent method (PSCM) \cite{GKKR96,MSK+95}, 
$U_{\rm c} = 1.46 W$, and to the result of the numerical renormalization-group 
(NRG) calculation, \cite{Bul99} $U_{\rm c} = 1.47 W$.
The linearized DMFT is able to predict very reliably the main trends in the 
dependence of $U_{\rm c}$ on the lattice geometry even for systems with
reduced translational symmetry. \cite{BP00,PN99a,PN99d,HB00}

The two-site DMFT is not restricted to the critical point $U=U_{\rm c}$ but 
more general. 
For $\mu=U/2$ and $n=1$ the self-energy of the two-site SIAM can be calculated 
analytically: \cite{Lan98} 
\begin{equation}
   \Sigma(\omega) = \frac{U}{2} + \frac{U^2}{8} \left( 
   \frac{1}{\omega-3V} + \frac{1}{\omega+3V}
   \right) \: .
\end{equation}
With Eq.\ (\ref{eq:zquasi}) this gives the quasi-particle weight:
\begin{equation}
   z = \frac{1}{1+U^2/36V^2} \: ,
\end{equation}
consistent with Eq.\ (\ref{eq:smallv}) for $V\mapsto 0$.
Together with the self-consistency condition (\ref{eq:cond2}) one arrives at:
\begin{equation}
   V^2 = \frac{M_2^{(0)}}{1+U^2/36V^2} \: .
\end{equation}
This algebraic but non-linear mean-field equation has the self-consistent 
solution
\begin{equation}
   V = \sqrt{M_2^{(0)} - \frac{U^2}{36}} 
\end{equation}
for $U<6\sqrt{M_2^{(0)}}=U_{\rm c}$ and $V=0$ else. 
This yields the $U$ dependence of the quasi-particle weight at half-filling:
\begin{equation}
   z = 1 - \frac{U^2}{U_{\rm c}^2} \: .
\end{equation}
The result is the same as in the Gutzwiller variational approach: \cite{BR70}
\begin{equation}
   z_{\rm BR} = 1 - \frac{U^2}{U_{\rm c, BR}^2} \: .
\end{equation}
However, the Brinkman-Rice critical interaction $U_{\rm c, BR} = -16 
\int_{-\infty}^0 dx \: x \: \rho_{0}(x) \approx 6.79 t^\ast > 6 t^\ast$ is 
considerably stronger.

\begin{figure}[t]
\begin{center}
\epsfig{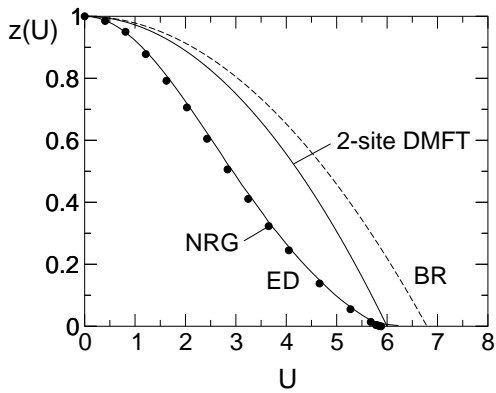}
\caption{
$U$ dependence of the quasi-particle weight at half-filling as obtained 
within the two-site DMFT, the Brinkman-Rice approach \cite{BR70} (dashed), 
the exact-diagonalization method \cite{CK94,Pot00} (solid), and the numerical 
renormalization-group approach \cite{Bul99} (dots).
} 
\label{fig:zofu}
\end{center}
\end{figure}

The result of the two-site DMFT is compared with the Brinkman-Rice solution 
and with the results of numerical solutions of the full DMFT by using the 
exact-diagonalization method \cite{CK94,Pot00} ($n_{\rm s}=8$ sites) and 
the NRG \cite{Bul99} in Fig.\ \ref{fig:zofu}.
There is a good agreement between the results of the ED and the NRG calculations
except for interactions close to $U_{\rm c}$ where the energy scale of the 
quasi-particle resonance cannot be resolved reliably within the ED method.
The two-site DMFT overestimates the quasi-particle weight in the whole 
$U$ range. However, the qualitative agreement with the full DMFT is better
than could be expected for the rather simple approach and clearly improves
on the result of the Gutzwiller method. 

The two-site DMFT interpolates between the trivial $U=0$ limit and 
$U=U_{\rm c}$.
For $U=U_{\rm c}$ it is a reasonable approximation to neglect (i) the 
influence of the Hubbard bands at high frequencies on the low-frequency 
(quasi-particle) peak and (ii) the internal structure of the quasi-particle 
peak (see the discussion in Ref.\ \onlinecite{BP00}). 
These are just the basic assumptions of the linearized DMFT which the
two-site approach reduces to for $U=U_{\rm c}$.
For $0 < U < U_{\rm c}$ these assumptions are less justified. 
Yet, the quadratic behavior of $z(U)$ for $U \mapsto 0$ as well as the
eventually linear behavior for $U \mapsto U_{\rm c}$ is consistent with 
the findings of the ED and NRG calculations. 

For $U>U_{\rm c}$ the self-consistent solution is given by $V=0$ which 
implies that the two-site DMFT reduces to the Hubbard-I approach in this 
case (Eq.\ (\ref{eq:sighub})).
This is a crude description of the Mott insulator, even if compared with
the Hubbard alloy-analogy solution and the iterative perturbation theory.
The main deficiency is that the widths of the Hubbard bands are largely
underestimated (see Ref.\ \onlinecite{GK93}, for example). 

\subsection{Fillings $n \ne 1$}

\begin{figure}[t]
\begin{center}
\epsfig{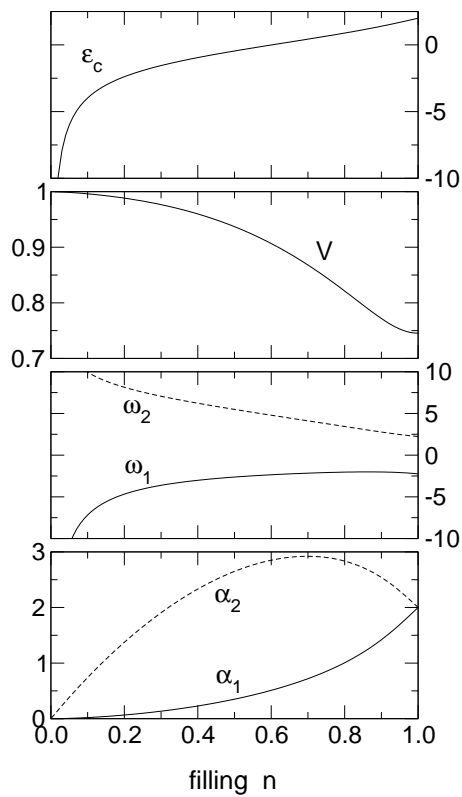}
\caption{
Filling dependence of the parameters of the effective two-site impurity
model $\epsilon_{\rm c}$ and $V$ and of the poles $\omega_1$, $\omega_2$ 
and respective weights $\alpha_1$, $\alpha_2$ of the self-energy
(see Eq.\ (\ref{eq:sigtwop})). $U=W=4$.
} 
\label{fig:siam}
\end{center}
\end{figure}

\begin{figure}[t]
\begin{center}
\epsfig{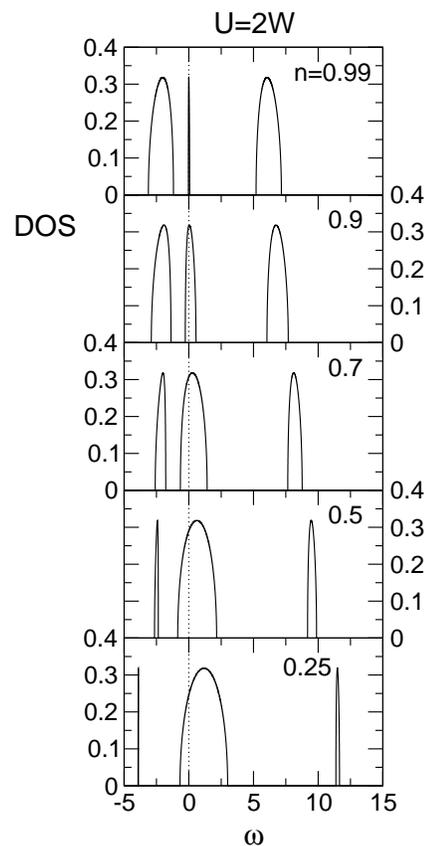}
\caption{
Density of states $\rho(\omega)$ for $U=2W=8$ and different fillings.
} 
\label{fig:dos}
\end{center}
\end{figure}

\begin{figure}[t]
\begin{center}
\epsfig{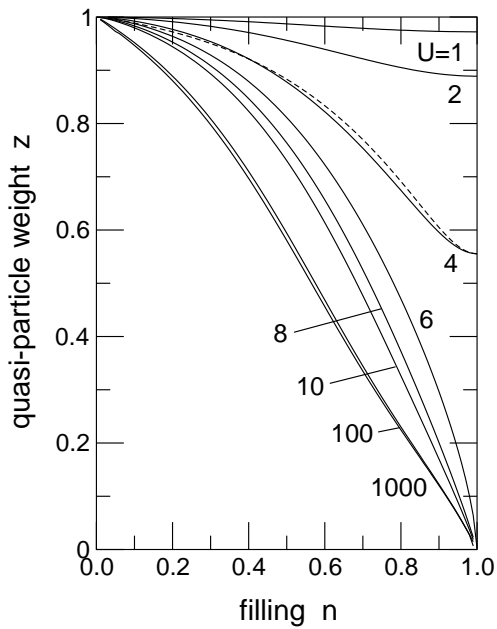}
\caption{
Quasi-particle weight as a function of the filling for different 
interaction strengths. Dashed line: calculation for $U=4$ with a
slightly modified theory (see text).
} 
\label{fig:zofn}
\end{center}
\end{figure}

\begin{figure}[t]
\begin{center}
\epsfig{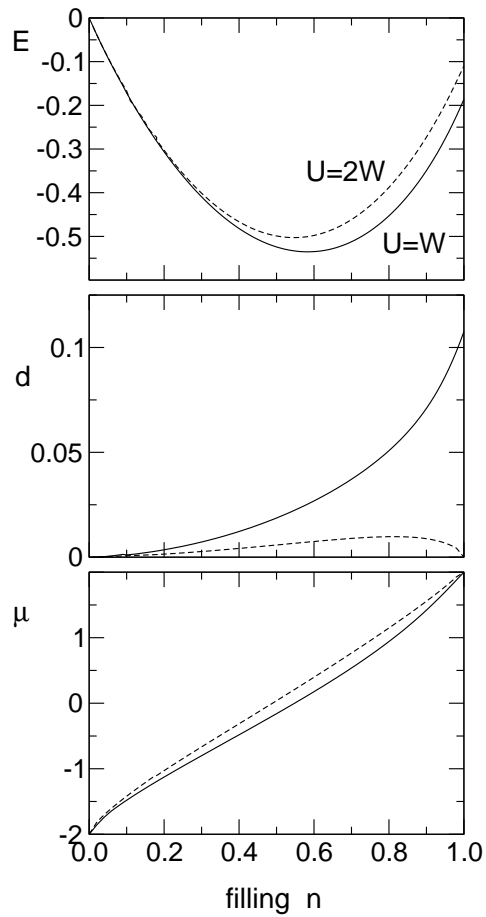}
\caption{
Filling dependence of the total energy $E$, the double occupancy
$d=\langle n_{\uparrow} n_{\downarrow} \rangle$, and the chemical
potential $\mu$. Results for $U=W$ (solid lines) and $U=2W$
(dashed lines).
} 
\label{fig:u4u8}
\end{center}
\end{figure}

For the symmetric case $n=1$ the first self-consistency equation $n=n_{\rm imp}$
is fulfilled trivially by particle-hole symmetry since $\epsilon_{\rm c} = U/2 = \mu$,
and only the hybridization strength $V$ has to be determined self-consistently.
For a thorough test of the two-site DMFT it is thus necessary to consider fillings
off half-filling, too. 

Fig.\ \ref{fig:siam} shows the self-consistent bath parameters $\epsilon_{\rm c}$
and $V$ as a function of the filling for $U=4$.
As $U$ is smaller than $U_{\rm c}$ the hybridization strength is finite for $n=1$.
For decreasing filling the system becomes less and less correlated, and consequently 
$V$ has to increase until $V=1=M_2^{(0)}$ for $n=0$. 
According to Eq.\ (\ref{eq:cond2}) this implies the correct value $z=1$ for the
quasi-particle weight in the empty-band limit.
At half-filling the one-particle energy of the bath site is given by 
$\epsilon_{\rm c}=2=U/2$. 
It decreases with decreasing filling $n$ and diverges on approaching the empty-band 
limit, $\epsilon_{\rm c} \mapsto -\infty$, as it is necessary to ensure a vanishing
occupancy of the impurity orbital $n_{\rm imp}=n \mapsto 0$ for finite $V$.
Note that both parameters, $\epsilon_{\rm c}$ and $V$, are smooth functions of the
filling.

Off half-filling the interacting impurity Green function continues to have four 
poles for $U>0$ and two poles for $U=0$.
This implies that the self-energy is a two-pole function of the form 
(\ref{eq:sigtwop}) not only for half-filling but also for $n \ne 1$.
The poles $\omega_1$ and $\omega_2$ of the self-energy and their respective weights 
$\alpha_1$ and $\alpha_2$ are shown in the lower panels of Fig.\ \ref{fig:siam}.
Again, it is noteworthy that these are smooth functions connecting the symmetric
point $n=1$ where $\omega_1+\omega_2=0$ and $\alpha_1=\alpha_2$ with the empty-band
limit where the poles become irrelevant since $\alpha_1,\alpha_2 \mapsto 0$.

It is well known \cite{GKKR96} that the density of states in the paramagnetic 
phase of the infinite-dimensional Hubbard model essentially consists of three 
peaks, the lower and the upper Hubbard band and a quasi-particle resonance in 
the vicinity of the Fermi energy.
An attractive feature of the two-site DMFT is that this general form of the 
density of states can be reproduced qualitatively. 
It is obvious that the two-pole structure of the self-energy results in a 
three-peak structure of the density of states (DOS)
$\rho(\omega) \equiv \rho_{0}(\omega + \mu - \Sigma(\omega))$.

Because the two-pole self-energy is real, the Hubbard bands and the 
quasi-particle resonance will be perfectly separated from each other on the 
real frequency axis.
Clearly, this is only a sketch of the true density of states -- the full DMFT
generally predicts the resonance to merge with one of the Hubbard 
peaks. \cite{JP93,GKKR96,KK96,PWN97}
The symmetric case at half-filling is an exception. 
For $U < U_{\rm c}$ but close to $U_{\rm c}$, a more or less clear separation 
of energy scales is found to be realized in fact. \cite{GKKR96,Bul99}

Results for the DOS are shown in Fig.\ \ref{fig:dos} for $U=2W=8$ at different
fillings.
Since $U > U_{\rm c}$ in this case, the Mott insulator is approached for 
$n\mapsto 1$.
As soon as there is a finite hole concentration $1-n$, a quasi-particle 
resonance appears the width of which becomes broader with decreasing filling.
The resonance is pinned to the Fermi energy.
For the symmetric case $n=1$ and $U<U_{\rm c}$ it has a maximum at $\omega=0$.
For decreasing $n<1$ the maximum shifts to a frequency $\omega_0>0$, and the 
asymmetry with respect to $\omega=0$ increases.
Concurrently, the upper Hubbard band shifts to higher frequencies while its
weight decreases.
All this is qualitatively correct when comparing with the full DMFT.
\cite{JP93,GKKR96,KK96}
However, the two-site DMFT largely underestimates the widths of the Hubbard bands 
and does not predict the quasi-particle resonance to merge with the lower Hubbard 
band not even for smaller fillings.
This is an obvious artifact.

The width of the resonance is determined by $z$.
The filling-dependence of the quasi-particle weight is shown in 
Fig.\ \ref{fig:zofn} for different $U$.
For $n \mapsto 0$ there is a linear trend of the quasi-particle weight 
$z-1 \propto n$.
For $n\mapsto 1$ the weight $z(n)$ behaves linearly when $U>U_{\rm c}$
and quadratically when $U<U_{\rm c}$.
Generally, $z$ is a monotonously decreasing function of the filling for $n<1$ 
and arbitrary $U$ and a monotonously decreasing function of $U$ for arbitrary 
$n$ which saturates in the limit $U \mapsto \infty$.
These results are very similar to those of the Gutzwiller approach 
\cite{Gut63,Gut65,VWA87} and qualitatively reproduce the results of the 
full DMFT \cite{Pot00} while quantitatively there are deviations similar 
as for the case $n=1$ which has been discussed already (see Fig.\ \ref{fig:zofu}).

The same holds for the filling dependence of the internal energy $E$, the 
double occupancy $d \equiv \langle n_\uparrow n_\downarrow \rangle$ and 
the chemical potential $\mu$ which is shown in Fig.\ \ref{fig:u4u8}.
For $U=W=4$ as well as for $U=2W=8$ the chemical potential monotonously
increases with increasing filling as it is required by thermodynamic stability.
From the equation of motion for the on-site lattice Green function one 
can derive simple expressions for the kinetic energy per site,
\begin{equation}
  E_{\rm kin} = 2 \int_{-\infty}^0 d\omega \:
  (\omega + \mu - \Sigma(\omega)) \rho(\omega) \: ,
\end{equation}
and for the potential energy per site:
\begin{equation}
  E_{\rm pot} = \int_{-\infty}^0 d\omega \:
  \Sigma(\omega) \rho(\omega) \: .
\end{equation}
The internal energy per site is given by $E=E_{\rm kin}+E_{\rm pot}$, and
the average number of doubly occupied sites by 
$d \equiv \langle n_{\uparrow} n_\downarrow \rangle = E_{\rm pot} / U$.
The double occupancy vanishes in the strong-coupling limit $U\mapsto \infty$.
For $U=W<U_{\rm c}$ it is a monotonously increasing function of $n$ while
$d$ is small (but non-zero) for $U=2W>U_{\rm c}$ at half-filling.

\subsection{Test of Fermi-liquid relations}

\begin{figure}[t]
\begin{center}
\epsfig{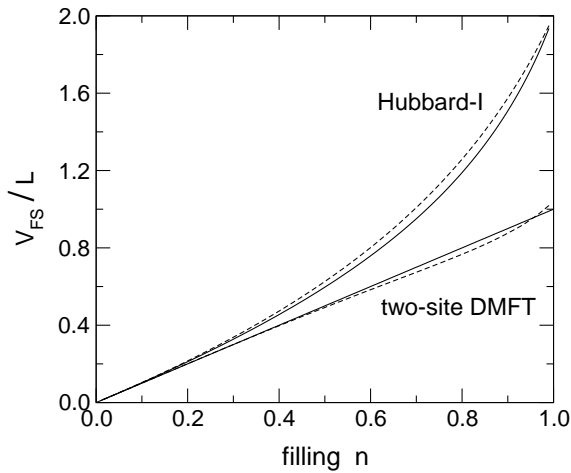}
\caption{
Normalized ``Fermi-surface volume'' $V_{\rm FS} / L$ as a function of the 
filling $n$ for $U=W$ (solid lines) and $U=2W$ (dashed lines).
Results as obtained within the two-site DMFT and the Hubbard-I 
approximation.
} 
\label{fig:lutt}
\end{center}
\end{figure}

\begin{figure}[t]
\begin{center}
\epsfig{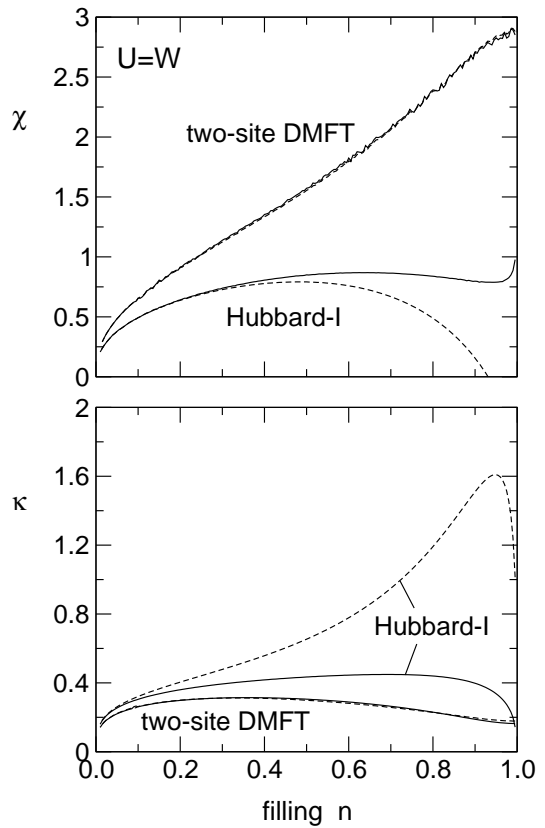}
\caption{
Charge susceptibility (compressibility) $\kappa$ and spin susceptibility $\chi$
as functions of the filling for $U=W=4$. 
Results for the two-site DMFT and the Hubbard-I approximation as indicated.
Solid lines: $\kappa$ and $\chi$ calculated from the definitions 
(\ref{eq:kappa1}) and (\ref{eq:chi1}).
Dashed lines: $\kappa$ and $\chi$ calculated from the Fermi-liquid representations
(\ref{eq:kappa2}) and (\ref{eq:chi2}).
} 
\label{fig:susz}
\end{center}
\end{figure}

\begin{figure}[t]
\begin{center}
\epsfig{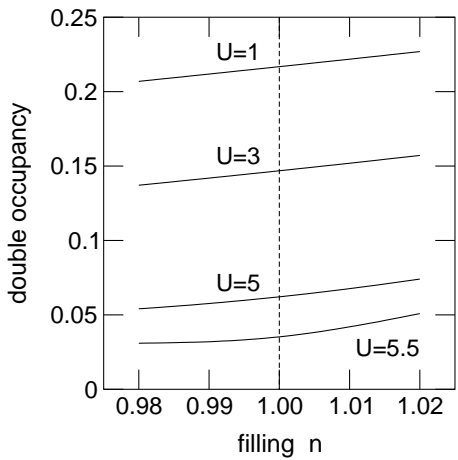}
\caption{
Filling dependence of the double occupancy 
$d=\langle n_\uparrow n_\downarrow \rangle$
close to half-filling for different $U$.
} 
\label{fig:thermo}
\end{center}
\end{figure}

Opposed to the full DMFT, the two-site DMFT is not a conserving approach in
the sense of Baym and Kadanoff. \cite{BK61}
Consequently, the two-site DMFT cannot be expected to respect certain exact
Fermi-liquid relations. \cite{LW60}
Drastic violations of the Luttinger theorem and Fermi-liquid relations for 
the charge and spin susceptibilities are well known \cite{BE95} for simple 
non-conserving theories such as the Hubbard-I approximation, \cite{Hub63}
the Roth two-pole approximation, \cite{Rot69} or the spectral-density 
approach, \cite{NB89} for example.
For the two-site DMFT it is shown here that there are violations of 
Fermi-liquid relations indeed.
Surprisingly, however, these are fairly small.

One important property of a Fermi-liquid is that the self-energy $\Sigma(\omega)$ 
is real at $\omega=0$, i.\ e.\ $\mbox{Im} \Sigma(i0^+) = 0$. 
This guarantees the existence of a Fermi surface which (for a Bravais lattice) is 
defined by the set of points in $\bf k$ space with $0=\mu-\epsilon({\bf k})-\Sigma(0)$
where $\epsilon({\bf k})$ is the free band dispersion.
The Luttinger sum rule \cite{LW60} states that the volume enclosed by the Fermi
surface,
\begin{equation}
  V_{\rm FS} = 2 \sum_{\bf k} \Theta(\mu - \epsilon({\bf k}) - \Sigma(0)) \: ,
\end{equation}
is equal to the average particle number:
\begin{equation}
  V_{\rm FS} = \langle N \rangle \: . 
\end{equation}
Here, $\Theta$ is the usual step function, 
and $\langle N \rangle = \langle \sum_{i\sigma} n_{i\sigma} \rangle = L n$
where $L$ is the number of lattice sites. 
The Fermi-surface volume can be calculated in the following way:
\begin{eqnarray}
  V_{\rm FS} / L &=& 2 \int_{-\infty}^\infty dx \: 
  \rho_{0}(x) \: \Theta(\mu - x - \Sigma(0)) 
\nonumber \\ 
  &=& 2 \int_{-\infty}^0 dx \: \rho_{0}(x+\mu-\Sigma(0)) 
\label{eq:vfs}
\end{eqnarray}
to be compared with the filling as given by Eq.\ (\ref{eq:fill}).
The equation $V_{\rm FS} / L = n$ with $V_{\rm FS} / L$ given by (\ref{eq:vfs})
is a reformulation of the Luttinger sum rule that is also valid for the Bethe
lattice.

In Fig.\ \ref{fig:lutt} the Fermi-surface volume $V_{\rm FS} / L$ is compared 
with the filling $n$ for different $U$.
For moderate interactions ($U=W$, solid line) the Luttinger theorem 
$V_{\rm FS} / L = n$ is almost exactly fulfilled in the whole range of 
fillings while deviations of a few percent are found for strong Coulomb 
interaction ($U=2W$, dashed line) near half-filling.
The results of the Hubbard-I approach are shown for comparison. 
It is seen that in this case the Fermi-surface volume is strongly 
overestimated up to a factor 2 near half-filling and irrespective of $U$.

An alternative formulation of the Luttinger theorem is given by \cite{MH89c} 
\begin{equation}
  \mu = \mu_0 + \Sigma(0) \: ,
\label{eq:luttmh}
\end{equation} 
where $\mu_0$ is the chemical potential for $U=0$.
Replacing the first self-consistency equation (\ref{eq:cond1}) by 
Eq.\ (\ref{eq:luttmh}), defines a variant of the two-site DMFT where
the Luttinger theorem is enforced.
The resulting filling dependence of the quasi-particle weight is 
shown in Fig.\ \ref{fig:zofn} for $U=4$ by the dashed line. 
If the original two-site DMFT respected the Luttinger theorem, there
would not be any difference compared with the result of the variant.
As is seen in Fig.\ \ref{fig:zofn}, the difference is non-zero but rather
small.

By means of perturbation theory to all orders \cite{LW60} the compressibility 
or charge susceptibility
\begin{equation}
   \kappa = \frac{\partial n}{\partial \mu} 
\label{eq:kappa1}
\end{equation}
can be shown to be related to the DOS and the self-energy at $\omega=0$:
\begin{equation}
   \kappa = 2 \rho(0) \left( 1 - \frac{\partial \Sigma(0)}{\partial \mu} \right) \: .
\label{eq:kappa2}
\end{equation}
Similarly, the spin susceptibility
\begin{equation}
   \chi = \frac{\partial m}{\partial b} \Big|_{b=0}
\label{eq:chi1}
\end{equation}
is given by the expression
\begin{equation}
   \chi = \rho(0) \left( 2 - 
   \frac{\partial (\Sigma_\uparrow(0) - \Sigma_\downarrow(0))}{\partial b} 
   \right) \: .
\label{eq:chi2}
\end{equation}
These equations provide two more exact Fermi-liquid relations.

To calculate the spin susceptibility according to Eq.\ (\ref{eq:chi1}),
the formalism has to be generalized for the spin-polarized case.
Consider a constant external magnetic field $b$ in $z$ direction which 
couples to the total spin:
\begin{equation}
  H = H_0 - 2 b S_z = H_0 - b \sum_{i\sigma} q_\sigma n_{i\sigma} \: ,
\label{eq:field}
\end{equation}
where $H_0 = H(b=0)$ is the Hamiltonian (\ref{eq:hubbard}), 
$n_{i\sigma} = c_{i\sigma}^\dagger c_{i\sigma}$, and $q_\uparrow=1$, 
$q_\downarrow=-1$ is a sign factor.
The field strength is given by $b$.
To account for a finite $b$ in the formalism 
it is sufficient to redefine the on-site hopping
$t_0 \mapsto t_{0\sigma} = t_0 - q_\sigma b$ and likewise
$\epsilon_{\rm d} \mapsto \epsilon_{\rm d\sigma} = \epsilon_{\rm d} - q_\sigma b$.
Furthermore, the bath parameters may become spin-dependent:
$\epsilon_{\rm c} \mapsto \epsilon_{\rm c \sigma}$ and
$V \mapsto V_\sigma$ in the impurity model (\ref{eq:imp}).
The self-consistency conditions (\ref{eq:cond1}) and (\ref{eq:cond2}) now read:
\begin{equation}
  n_{{\rm d} \sigma} \equiv \langle d^\dagger_\sigma d_\sigma \rangle 
  \stackrel{!}{=} n_\sigma \equiv \langle n_{i\sigma} \rangle 
  \: , \hspace{5mm}
  V^2_\sigma \stackrel{!}{=} z_\sigma M_2^{(0)} \: .
\label{eq:condspin}
\end{equation}
The magnetization is given by $m=n_\uparrow - n_\downarrow$.

Fig.\ \ref{fig:susz} shows the filling dependence of $\kappa$ and $\chi$ as 
calculated from Eqs.\ (\ref{eq:kappa1}) and (\ref{eq:chi1}) to be compared 
with $\kappa$ and $\chi$ as calculated from Eqs.\ (\ref{eq:kappa2}) and 
(\ref{eq:chi2}).
For $U=4$ there are hardly any differences.
Small differences of the order of a few percent are found for $U=8$ (not
shown).
On the other hand, within the Hubbard-I approach the filling-dependence of 
$\kappa$ and $\chi$ strongly depends on the way it is calculated.
The respective results are shown in Fig.\ \ref{fig:susz} for comparison. 

Concluding, one can state that the Luttinger theorem as well as certain
Fermi-liquid relations are well respected by the two-site DMFT.
The same holds for the question of thermodynamic consistency:
Although the two-site DMFT cannot be derived from an explicit thermodynamic 
potential, consistency relations such as 
\begin{equation}
  \frac{\partial^2 E}{\partial U \partial n} =
  \frac{\partial \mu}{\partial U} \Big|_n = 
  \frac{\partial \langle n_{\uparrow} n_{\downarrow} \rangle}{\partial n}
  \Big|_U
\end{equation}
are found to hold to a comparatively good approximation.
An example is shown in Fig.\ \ref{fig:thermo}. 
At half-filling the chemical potential is given by $\mu=U/2$. 
Therefore, as a function of the filling the double occupancy 
$\langle n_{\uparrow} n_{\downarrow} \rangle$ should have the slope 
$\partial \langle n_{\uparrow} n_{\downarrow} \rangle/ \partial n = 1/2$
at $n=1$ for any (fixed) $U$.
To a very good approximation this is reproduced by the results in 
Fig.\ \ref{fig:thermo}.

\section{Conclusions}
\label{sec:conclusion}

The two-site DMFT can be characterized as an approximate DMFT scheme which 
refers to a single-impurity Anderson model consisting of two sites only, one 
impurity and one bath site.
Obviously, with a two-site SIAM the original self-consistency condition of
the full DMFT can no longer be fulfilled, and consequently a comparatively 
crude approximation has to be tolerated.
The idea, however, is to construct in this way the most simple approach that 
keeps the essence of the DMFT, namely the mapping onto an effective impurity 
model the bath parameters of which are determined self-consistently.
In fact, for the single-band Hubbard model some of the calculations can be 
performed even analytically and there are no serious problems to be expected 
for a numerical treatment of multi-band models.

Any realization of the DMFT requires a repeated solution of the impurity 
model which itself poses a complex many-body problem.
An exact and unproblematic solution is possible for a SIAM with a finite, 
small number of degrees of freedom -- at the expense of an approximate mapping. 
This idea of the two-site DMFT is similar in spirit to the that of the 
exact-diagonalization method which aims at a minimization of the errors 
due to the discretization of the hybridization function by including as 
many sites $n_{\rm s}$ as feasible numerically.
Contrary, the two-site DMFT stays with the case $n_{\rm s}=2$ and can thus 
be considered as a ``two-site ED'' method. 
The latter, however, is not unique as it depends on the fit procedure used
for the numerical determination of the bath parameters. \cite{CK94,SRKR94}
The advantage of the two-site DMFT is that it is based on well motivated 
self-consistency conditions to fix the bath.

In principle, an extension of the two-site DMFT towards an $n_{\rm s}$-site
DMFT is conceivable for the single-band Hubbard model. 
This requires the consideration of higher-order terms in the high- and 
low-frequency expansion of the original self-consistency equation to fix 
the additional $2n_{\rm s}-4$ bath parameters.
The computational effort, however, grows exponentially with increasing
$n_{\rm s}$.
Furthermore, higher-order static correlation functions would appear in the
$1/\omega$ expansion of the on-site lattice Green function (\ref{eq:greenhigh}).
At the order $1/\omega^4$ the correlated-hopping correlation function
$B_\sigma \propto \sum_{i \ne j} t_{ij} \langle c_{i\sigma}^\dagger c_{j\sigma} 
(2 n_{i-\sigma} -1) \rangle$ can be expressed rigorously in terms of $G(\omega)$
by a sum rule \cite{PHWN98} similar to Eq.\ (\ref{eq:n}).
At still higher orders, however, additional approximations must be tolerated.
 
The implementation of the two-site DMFT is straightforward, and the numerics 
is stable for the entire parameter space.
It has to be stressed that the approach allows for computations that are faster
by several orders of magnitude compared to numerically exact approaches such 
as the QMC or ED method.
This advantage qualifies the approach for comprehensive investigations as are
necessary e.\ g.\ for the determination of phase diagrams. 
In this respect it is comparable with the iterative perturbation theory (IPT),
\cite{GKKR96,KK96,PWN97} for example.

The single-band Hubbard model has been considered here to illustrate the 
construction of the theory and to show up its advantages and limitations. 
Obviously, for the mere single-band model, the two-site DMFT cannot really 
compete against other more suitable methods. 
The main field of application are multi-band Hubbard-type models which 
require the solution of more complicated effective impurity problems, or 
lattice models with reduced translational symmetry where one has to solve
different impurity problems simultaneously. \cite{PN99a,PN99d}
Opposed to the IPT, \cite{LK98} for example, the extension of the two-site 
DMFT to the multi-band case is straightforward.
Still the computational effort will be extremely small as compared 
to the numerically exact approaches.

Taking the mass-enhancement factor $m^\ast/m = z^{-1}$ as a measure for 
the strength of correlation effects, the Mott-transition point $n=1$ and
$U=U_{\rm c}$ obviously has a distinguished position in the phase diagram.
Here the two-site DMFT reduces to the previously developed linearized DMFT
\cite{BP00} which rather accurately predicts (the mean-field) $U_{\rm c}$ 
for the Hubbard model on different lattices \cite{BP00,PN99a,PN99d} and 
also for multi-band models. \cite{HB00,OBH00}
The two-site DMFT can be considered as an extension of the linearized DMFT
to the entire parameter space. 
It interpolates between the Mott-transition point where $z=0$ and the 
uncorrelated limit $U=0$ or $n=0$ where $z=1$.
Similar as the linearized theory, however, it cannot be controlled by a 
``small parameter'' but is based on a physically motivated approximation.

The two-site DMFT is the most simple approach that describes the transition 
from the Mott insulating state to the Fermi liquid as a bifurcation 
scenario: \cite{GKKR96}
At half-filling and for $U>U_{\rm c}$ the approach reduces to the Hubbard-I 
approximation and yields the Mott-insulating solution.
At $U=U_{\rm c}$ a metallic solution splits off from the insulating one, the 
former being stable for $U<U_{\rm c}$.
Below $U_{\rm c}$ the two-site DMFT predicts a Fermi-liquid state with 
$z=1-U^2/U_{\rm c}^2$. 
This is the same as the Brinkman-Rice-Gutzwiller result \cite{BR70} (albeit 
with a different $U_{\rm c}$).
Although conceptually the two-site DMFT is quite different from the Gutzwiller
method, \cite{Gut63,Gut65} the results for dependence of $z$ on $U$ and $n$ 
are very similar.
In particular, both approaches yield $z(\delta)-z(0) \propto \delta^2$ for
$U<U_{\rm c}$ and $z(\delta)-z(0) \propto \delta$ for $U>U_{\rm c}$ in the 
limit $\delta=1-n \mapsto 0$ (see Fig.\ \ref{fig:zofn} and Ref.\ \onlinecite{VWA87}).

Similar as the IPT but opposed to the Gutzwiller method, the two-site DMFT does 
not respect the Luttinger sum rule for the invariance of the Fermi-surface volume
since it is not a conserving theory in the sense of Baym and Kadanoff \cite{BK61}
and cannot be derived from an explicit expression for a thermodynamic potential.
On the other hand, it is remarkable that the deviations from the Luttinger sum 
rule and also the deviations from different exact Fermi-liquid and thermodynamical 
consistency relations are rather small.
The comparison with the Hubbard-I approximation (which drastically violates these
relations) suggests that this is due to the appearance of the quasi-particle peak
in the spectral function.
The Hubbard-I approximation can be considered as a local impurity approximation:
The exact self-energy of the single-site (atomic) model is taken as an approximation 
for the self-energy of the lattice model.
This approach and similar but improved theories \cite{Rot69,NB89} yield high-frequency 
excitations such as the Hubbard bands but fail to reproduce Fermi-liquid properties 
at low frequencies.
To get the (low-frequency) quasi-particle peak in addition and to restore the 
Fermi-liquid physics qualitatively, it is sufficient to couple merely a single 
(bath) degree of freedom to the single-site model.
This also holds for the multi-band case.
A fully conserving and consistent theory, however, can only be obtained by
coupling to an infinite number of bath sites as in the full DMFT.

Concluding, the two-site DMFT is a simple but non-perturbative mean-field 
approach to correlated lattice models.
Compared to the full DMFT and to exact Fermi-liquid and thermodynamic 
relations, it yields satisfactory results for the Mott transition and 
the Fermi-liquid phase in the single-band Hubbard model with a minimum 
computational effort.
The quality of the method, when applied to different physical problems, 
it not clear a priori and has to be examined. 
In particular, a comparison between the two-site and the full DMFT concerning 
magnetic order in the single-band model as well as the question of finite 
temperatures are intended for the future.
Primarily, future applications shall address the manifestly complex phase 
diagrams for spin, charge and orbital order in multi-band Hubbard-type models.
Here the two-site DMFT may serve to give a quick and comprehensive though rough 
overview of the main physics which can complement more thorough but selective 
studies.

\acknowledgments

This work is supported by the Deutsche Forschungsgemeinschaft within the 
Sonderforschungsbereich 290.

\appendix

\section{Multi-band models}
\label{sec:multi}

The most general multi-band model with on-site interaction reads:
\begin{eqnarray}
  H & = & \sum_{i_1i_2\alpha_1\alpha_2\sigma} t_{i_1i_2\alpha_1\alpha_2} \:
  c^\dagger_{i_1\alpha_1\sigma} c_{i_2\alpha_2\sigma} 
\nonumber \\
  & + & \frac{1}{2} 
  \sum_{i\sigma\sigma'} \sum_{\alpha_1\alpha_2\alpha_3\alpha_4} 
  U_{\alpha_1\alpha_2\alpha_4\alpha_3} \:
  c^\dagger_{i\alpha_1\sigma} 
  c^\dagger_{i\alpha_2\sigma'}
  c_{i\alpha_3\sigma'} 
  c_{i\alpha_4\sigma} \: .
\nonumber \\
\label{eq:mhubbard}
\end{eqnarray}
In most cases it is sufficient to take into account interaction parameters 
that are labeled by two indices, the direct interaction 
$U_{\alpha\alpha'}=U_{\alpha\alpha'\alpha\alpha'}$ and the
exchange interactions $J_{\alpha\alpha'}=U_{\alpha\alpha'\alpha'\alpha}$ and 
$J'_{\alpha\alpha'}=U_{\alpha\alpha\alpha'\alpha'}$.
Here $\alpha=1,...,r$ is an orbital index, and $r$ denotes the number of orbitals.
Fourier transformation of the hopping $t_{i i'\alpha\alpha'}$ to ${\bf k}$ 
space yields the free Hamilton matrix $t_{\alpha\alpha'}({\bf k})$ the
eigenvalues of which represent the free band structure: $\epsilon_m({\bf k})$
where $m=1,...,r$ is the band index.
The orbitally resolved free density of states is given by:
\begin{equation}
  \rho_{0,\alpha}(\omega) = \frac{1}{L} \sum_{{\bf k}m}
  |\phi_{\alpha m}({\bf k})|^2 \delta(\omega - \epsilon_m({\bf k})) \: ,
\end{equation}
where $\phi_{\alpha m}({\bf k})$ is the $\alpha$-th component of the $m$-th
eigenstate at each $\bf k$, namely $\sum_{\alpha \alpha'}
\phi_{\alpha m}({\bf k})^\ast t_{\alpha\alpha'}({\bf k}) 
\phi_{\alpha' m'}({\bf k}) = \epsilon_m({\bf k}) \delta_{mm'}$.

Assuming the self-energy to be local, it is easy to see from the usual diagram 
expansion that it must be diagonal with respect to the orbital index: 
$\Sigma_{\alpha\alpha'\sigma}(\omega) = \delta_{\alpha\alpha'} 
\Sigma_{\alpha\sigma}(\omega)$.
Consider a cubic lattice and let $\alpha$ refer to orbitals with an angular 
dependence given by the cubic harmonics. 
The lattice symmetries then require the on-site ($i=i'$) elements of the 
lattice Green 
function $G_{ii'\alpha\alpha'\sigma}(\omega) = \langle \langle 
c_{i\alpha\sigma} ; c_{i'\alpha'\sigma}^\dagger \rangle \rangle_\omega$ to be
diagonal with respect to $\alpha$: 
\begin{equation}
  G_{ii \alpha\alpha'\sigma}(\omega) = \delta_{\alpha\alpha'} 
  G_{\alpha\sigma}(\omega) \: .
\end{equation}
Using the Dyson equation, the on-site Green function can generally be written as
\begin{equation}
  G_{\alpha\sigma}(\omega) = \frac{1}{L} \sum_{{\bf k}}
  \left[
  \frac{1}{\omega + \mu - {\bf t}({\bf k}) - {\bf \Sigma}_\sigma(\omega)} 
  \right]_{\alpha\alpha}
  \: .
\label{eq:mgreen}
\end{equation}
Here ${\bf \Sigma}_\sigma(\omega)$ is the ($\bf k$-independent) diagonal $r \times r$ 
self-energy matrix and ${\bf t}({\bf k})$ the $\bf k$-dependent (non-diagonal) 
free Hamilton matrix with the elements $t_{\alpha\alpha'}({\bf k})$.

Within the DMFT the model (\ref{eq:mhubbard}) can be mapped onto the following
impurity model:
\begin{eqnarray} 
  H_{\rm imp} 
  & = & 
  \sum_{\alpha\sigma} \epsilon_{{\rm d}\alpha} \:
  d^\dagger_{\alpha\sigma} d_{\alpha\sigma} 
\nonumber \\
  & + & \frac{1}{2} 
  \sum_{\sigma\sigma'} \sum_{\alpha_1\alpha_2\alpha_3\alpha_4} 
  U_{\alpha_1\alpha_2\alpha_4\alpha_3} \:
  d^\dagger_{\alpha_1\sigma} 
  d^\dagger_{\alpha_2\sigma'}
  d_{\alpha_3\sigma'} 
  d_{\alpha_4\sigma} 
\nonumber \\
  & + & \sum_{\alpha \sigma, k=2}^{n_{\rm s}} 
  \epsilon_{k\alpha\sigma} \: 
  a^\dagger_{k\alpha\sigma} a_{k\alpha\sigma}
\nonumber \\
  & + & 
  \sum_{\alpha\sigma, k=2}^{n_{\rm s}}
  V_{k\alpha\sigma} \: 
  (d^\dagger_{\alpha\sigma} a_{k\alpha\sigma} + \mbox{h.c.}) \: ,
\label{eq:mimp} 
\end{eqnarray}
with $\epsilon_{{\rm d}\alpha} = t_{0\alpha} = t_{ii\alpha\alpha}$. 
The impurity Green function 
$G_{{\rm imp},\alpha\sigma}(\omega) = \langle \langle d_{\alpha\sigma} ; 
d_{\alpha\sigma}^\dagger \rangle \rangle_{\omega}$ is given by
\begin{equation}
  G_{{\rm imp},\alpha\sigma}(\omega) =
  \frac{1}{\omega + \mu - \epsilon_{{\rm d}\alpha}
  - {\Delta}_{\alpha\sigma}(\omega) - {\Sigma}_{{\rm imp},\alpha\sigma}(\omega)} 
  \: .
\label{eq:mimpgreen}
\end{equation}
The hybridization function $\Delta_{\alpha\sigma}(\omega) = \sum_k 
V_{k\alpha\sigma}^2 / (\omega + \mu - \epsilon_{k\alpha\sigma})$ 
and the impurity self-energy are diagonal with respect to $\alpha$.

The lattice self-energy can be derived by functional differentiation from the
Luttinger-Ward functional, $T {\Sigma}_{i'i\alpha'\alpha\sigma}(i\omega) = \delta 
\Phi / \delta {G}_{ii'\alpha\alpha'\sigma}(i\omega)$. \cite{LW60} 
For any finite-dimensional lattice the DMFT consists in the assumption that
the self-energy be local.
This implies that $\Phi$ depends on the on-site propagator 
$G_{\alpha\sigma}(\omega)$ only.
Hence, the functional $\Phi$ and thus the functional ${\cal S} = \delta \Phi /
\delta G$ are the same for both, the lattice model ($\Sigma = {\cal S}[G]$) and 
the impurity model ($\Sigma_{\rm imp}= {\cal S}[G_{\rm imp}]$).
One can proceed as for the single-band case: 
If the bath parameters $\epsilon_{k\alpha\sigma}$ and $V_{k\alpha\sigma}$ 
are chosen such that 
\begin{equation}
  \Delta_{\alpha\sigma}(\omega) = \omega + \mu - \epsilon_{{\rm d} \alpha} - 
    \Sigma_{{\rm imp},\alpha\sigma}(\omega) -
    \frac{1}{G_{\alpha\sigma}(\omega)} \: ,
\label{eq:mhybsc}  
\end{equation}
i.\ e.\ such that the DMFT self-consistency condition,
\begin{equation}
   G_{{\rm imp},\alpha\sigma}(\omega) \stackrel{!}{=} G_{\alpha\sigma}(\omega) \: ,
\label{eq:msc}
\end{equation}
is fulfilled, then at once $\Sigma_{{\rm imp},\alpha\sigma}(\omega) = 
\Sigma_{\alpha\sigma}(\omega)$, and the usual self-consistent procedure 
can be set up.

If different orbitals are equivalent due to lattice symmetries as, for 
example, the three $t_{2g}$ orbitals in a cubic $d$-band system, one can make 
use of some simplifications:
The self-energy $\Sigma_{\alpha\sigma}(\omega) = \Sigma_\sigma(\omega)$ and 
the DOS $\rho_{0,\alpha}(\omega) = \rho_{0}(\omega)$ are independent of 
the orbital index, and the on-site Green function is simply given by:
\begin{equation}
  G_{\alpha\sigma}(\omega) = G_\sigma(\omega) = \int_{-\infty}^\infty dx \:
  \frac{\rho_{0}(x)}{\omega + \mu - x - \Sigma_\sigma(\omega)} \: .
\end{equation}
The lattice model can be mapped onto a simpler impurity model with 
$\epsilon_{k\alpha\sigma} = \epsilon_{k\sigma}$ and
$V_{k\alpha\sigma} = V_{k\sigma}$.
In this case the hybridization function and the impurity self-energy are 
$\alpha$-independent, consistent with Eq.\ (\ref{eq:mhybsc}).

The two-site DMFT is constructed straightforwardly.
The number of (two-fold spin-degenerate) one-particle orbitals in the impurity
model is $M = r + r (n_{\rm s}-1) = r n_{\rm s}$.
For $n_{\rm s}=2$ and for the general case of $r$ non-equivalent orbitals 
there are (for each spin direction) $2r$ bath parameters to be determined, 
the one-particle energies $\epsilon_{\alpha\sigma}$ and the hybridization 
strengths $V_{\alpha\sigma}$ for $\alpha=1,...,r$.
The comparison of the high-frequency expansions of $\Sigma_{\alpha\sigma}(\omega)$ 
and $\Sigma_{{\rm imp},\alpha\sigma}(\omega)$ to lowest order leads to a first set 
of $r$ self-consistency conditions:
\begin{equation}
  n_{{\rm d}, \alpha \sigma} \equiv 
  \langle d^\dagger_{\alpha\sigma} d_{\alpha\sigma} \rangle 
  \stackrel{!}{=} n_{\alpha\sigma} \equiv 
  \langle c^\dagger_{i\alpha\sigma} c_{i\alpha\sigma} \rangle 
  \: .
\label{eq:mc1}
\end{equation}
with $n_{\alpha\sigma} = (-1/\pi) \int_{-\infty}^0 \mbox{Im} \: 
G_{\alpha\sigma}(\omega+i0^+) \: d\omega$.

In the low-frequency limit the self-energy is expanded as 
$\Sigma_{\alpha\sigma}(\omega) = a_{\alpha\sigma} + (1-z_{\alpha\sigma}^{-1}) 
\omega + {\cal O}(\omega^2)$ where $z_{\alpha\sigma}$ is the orbital-dependent 
quasi-particle weight.
Inserting the self-energy expansion up to the linear order into Eqs.\
(\ref{eq:mgreen}) and (\ref{eq:mimpgreen}), yields the respective coherent
parts of the on-site lattice and the impurity Green function.
Analogous to the single-band case one gets:
\begin{eqnarray}
  G_{{\rm imp},\alpha\sigma}^{\rm (coh)}(\omega) 
  &=& \frac{z_{\alpha\sigma}}{\omega} + \frac{z_{\alpha\sigma}^2 
  (\epsilon_{{\rm d}\alpha} - \mu + a_{\alpha\sigma})}{\omega^2} 
\nonumber \\
  &+& \frac{z_{\alpha\sigma}^2 V_{\alpha\sigma}^2 
  + z_{\alpha\sigma}^3 
  (\epsilon_{{\rm d}\alpha} - \mu + a_{\alpha\sigma})^2}{\omega^3} 
\nonumber \\  
  &+& {\cal O}(1/\omega^4) \: .
\label{eq:mgreenimpcohhigh}
\end{eqnarray}
The expansion of the coherent on-site lattice Green function, however, is
slightly different:
\begin{eqnarray}
  G_{\alpha\sigma}^{\rm (coh)}(\omega) 
  &=& \frac{z_{\alpha\sigma}}{\omega} + \frac{z_{\alpha\sigma}^2 
  (t_{0\alpha} - \mu + a_{\alpha\sigma})}{\omega^2} 
\nonumber \\
  &+& \frac{z_{\alpha\sigma}^3 (t_{0\alpha} - \mu + a_{\alpha\sigma})^2}{\omega^3} 
\nonumber \\
  &+& z^2_{\alpha\sigma} \sum_{i'\alpha'} t_{ii'\alpha\alpha'} 
  z_{\alpha'\sigma} t_{i'i\alpha'\alpha} \frac{1}{\omega^3}
\nonumber \\
  &+& {\cal O}(1/\omega^4) \: .
\label{eq:mgreencohhigh}
\end{eqnarray}
From the comparison one obtains the second set of $r$ self-consistency conditions:
\begin{equation}
   V_{\alpha\sigma}^2 \stackrel{!}{=} \sum_{i'\alpha'} z_{\alpha'\sigma}
   t_{ii'\alpha\alpha'}^2 \: .
\label{eq:mc2}
\end{equation}
This includes an additional coupling of the different orbitals.
Both self-consistency conditions (\ref{eq:mc1}) and (\ref{eq:mc2}) reduce to 
Eq.\ (\ref{eq:condspin}) for the single-band case.

\end{document}